\documentclass[11pt]{article}
\usepackage{vietnam,epsfig}

\def\Journal#1#2#3#4{{#1} {\bf #2}, #3 (#4)}

\def\NPB{{\em Nucl. Phys.} B}
\def\PLB{{\em Phys. Lett.}  B}
\def\PRL{\em Phys. Rev. Lett.}

\def\be{\begin{equation}}
\def\ee{\end{equation}}
\def\bea{\begin{eqnarray}}
\def\eea{\end{eqnarray}}

\begin{document}
\vspace*{4cm}
\title{THE ORIGIN OF SPACETIME DIMENSIONALITY}

\author{ M. SAKELLARIADOU }

\address{Department of Physics, King's College, University of
London,\\ Strand, London WC2R 2LS, United Kingdom\\
Mairi.Sakellariadou@kcl.ac.uk}

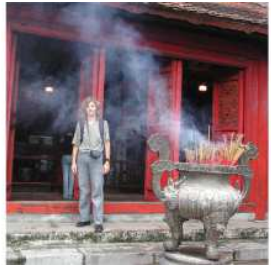
\maketitle \abstracts{I address the issue of spacetime dimensionality
within Kaluza-Klein theories and theories with large extra
dimensions. I review the arguments  explaining
the dimensionality of the universe, within
the framework of string gas cosmology and braneworld
cosmology, respectively.}

\section{Introduction}

Our universe is undoubtedly three-dimensional, and you may wonder why.  By
using a weak form of the anthropic principle, you may constrain the
number of spatial dimensions to be at least equal to 3; there is no
convincing argument however why it is just equal to 3, and not any
other bigger, even huge, number. Clearly there is no fundamental
physical principle to single out 3 spatial dimensions, and yet three
dimensions of space have a special status.

I address the {\sl puzzle} of spacetime dimensionality in the
framework of string theories. I assume a ten-dimensional
superstring theory, in the form of a nine-dimensional spatial torus with
time being the tenth dimension. The origin of spacetime dimensionality
can be equivalently stated as the mechanism through which the ten
dimensions required by string theory are reduced to the four
spacetime dimensions of our universe. This issue was first addressed
in the framework of Kaluza-Klein theories, while more recently it was
studied in the framework of braneworld models.  In what follows, I
briefly discuss the mechanism proposed to explain the origin of
spacetime dimensionality, first in the string gas scenario within the
Kaluza-Klein approach, and then in the framework of large extra
dimensions within IIB string theory.

\section{Kaluza-Klein approach: String gas scenario}
Within Kaluza-Klein theories the 6 extra spatial dimensions are rolled
up in a Calabi-Yau manifold with a size given by the string scale
$\sqrt{\alpha'}$.  In the context of string theory it holds
$T$-duality, a symmetry which is unique to string physics (it does not
hold for point particles because they do not have winding modes).  The
most notable example of $T$-duality is the {\sl target-space duality}
$R\rightarrow \alpha'/R$, which relates string theories compactified
on large and small tori by interchanging winding and Kaluza-Klein
states. Applying $T$-duality in cosmology led to the string gas
scenario~\cite{bv}, with distance having a different interpretation in
the two dual regimes: At a large radius the position coordinate is the
conjugate variable to momentum, $p=n/R$ as usual; at distances
smaller that the self-dual radius, one should use the dual coordinate
instead, i.e. the conjugate variable to winding $w=mR$ (where $n, m$ denote
the momentum and winding charges associated with the extra
dimensions). There is a minimum distance in string theory, with
distances smaller than the string scale being equivalent to large scales,
and under the hypothesis that the universe is a product of
circles the initial singularity is eliminated.  

According to the string gas scenario one can sketch the explanation
for the origin of the spacetime dimensionality as follows: The
universe starts with all spatial dimensions of the string
size. Winding modes prevent the corresponding dimensions from
expanding. Winding modes in general annihilate with anti-winding
modes, but they will miss each other in a ten-dimensional space. Only
in a four-dimensional hypersurface the worldsheets of winding and
anti-winding modes can naturally overlap, leading to annihilation and
thus the subsequent expansion of 3 spatial dimensions;
the winding modes prevent 6 spatial dimensions from expanding.

In conclusion, cosmology with string gases uses a new symmetry
($T$-duality) as well as new degrees of freedom (string winding
modes). It assumes homogeneous fields, adiabatic
approximation, weak coupling and toroidal spatial dimensions. Among
these assumptions, the adiabatic approximation and the weak coupling
regime are very restrictive from the string theory perspective. The
string gas scenario has been supported by cosmic string
experiments on a lattice~\cite{ms}, an approach which is justified
since we are dealing with classical aspects of strings.

Even though the string gas scenario, which has been later modified to
include branes, is quite appealing, it faces some important
challenges. The interaction probability relies on the dilaton, thus as
the dilaton runs to the weak coupling regime the interaction
probability goes to zero.  Moreover, viewing interactions as
intersections is an entirely classical argument. Finally, curvature
corrections and strong coupling behaviour could change the 
conclusions.

\section{Braneworld approach}
In the braneworld approach our universe is a three-dimensional brane
embedded in a higher dimensional bulk. Open strings end on the branes,
whereas closed strings move in the bulk.

In the context of the braneworld approach, it has been recently
proposed a scenario which could explain the survival of branes with
dimensionality at most equal to 3; one of such branes could become our
own universe. In what follows I highlight the main points of this
scenario. 

Consider a uniform distribution of Dirichlet branes of
dimensionality $p$, which are called $Dp$-branes. Working in a IIB
string theory, we thus have a uniform distribution of branes of odd
dimensionality. These branes are embedded in a bulk with spacetime
dimension denoted by $d$.  Assuming there are no brane
interactions at macroscopic distances, the interaction probability
should only depend on the relation between $p,d$ dimensions.
Considering at first interactions only among branes of the same
dimensionality, it has been argued~\cite{dks} that two $Dp$-branes
embedded in a $d$-dimensional spacetime generically intersect if
\be \label{bi} 2p+1\geq d-1~.  \ee For $d=10$ the higher brane
dimensionality for which intersection is avoided is equal to 3; all
higher dimensionality branes  eventually collide.  This condition
is however not enough to explain the origin of the spacetime
dimensionality. To claim so, one has to show that intersecting branes
are unstable, so that they eventually evaporate leaving behind $D3$-
and $D1$-branes (which are known as $D$-strings).

The evaporation of all branes having dimensionality at least equal to
4 has been shown~\cite{dks} under the following four (reasonable)
hypotheses: (i) The $d-1=9$ bulk coordinates are compactified on a
torus. Closed branes which do not wind around the torus shrink and
disappear, emitting closed string modes. (ii) If the intersection
process between two $Dp$-branes takes place on a hypersurface of
dimension $p-1$, the branes exchange partners and reconnect, reducing
their winding number until eventually they completely unwind and
subsequently evaporate. (iii) If intersection between two $Dp$-branes
takes place on a hypersurface of dimension smaller than $p-1$, the
branes  try to align (or anti-align) the directions with the
smallest opening angle, until the intersection manifold
reaches dimensionality $p-1$. Then, the two $Dp$-branes can reconnect
and subsequently evaporate. Finally, (iv) the total winding number of
all branes of a given dimensionality must be vanishing.

As it was argued~\cite{dks}, the hypotheses (i) and (ii) are quite
natural ones. More precisely, (i) is favoured from entropy
considerations and (ii) has been verified~\cite{ms} numerically for
one-dimensional topological defects, namely cosmic strings. The only
difference is that in the case of brane interactions the intersection
probability ${\cal P}$ can be considerably lower than 1;
${\cal P}=1$ for one-dimensional topological defects (cosmic
strings). A lower intersection probability may just slow down the whole
process of brane interactions. The validity of the result found for
string intersections, in the case of higher (than 1)
dimensionality branes can be explained by applying T-duality.  More
precisely, if two $Dp$-branes intersect in a manifold of
dimensionality $p-1$, then by applying $T$-duality in the $p-1$ common
directions we reduce it to the case of two one-dimensional branes
intersecting in a point. The last hypothesis, (iv), is not really
essential for our argument. One can therefore relax it, in which case
some higher dimensionality (at least $p=4$) branes can remain, but
even then they will be just a few as compared to the $D3$-branes.

The crucial hypothesis is (iii) and its validity was explicitly
shown~\cite{dks}. I will briefly sketch the proof. The brane
interactions can be described by an interaction potential, $V$, derived
from the scattering amplitude of open strings ending on the
branes. The force between the branes is the gradient of $V$. Consider
two $D4$-branes initially extended in the $(2,4,6,8)$-directions and
separated by a distance $y$ in 1-direction. Rotate one of them by
angle $\phi_1$ in the $(2,3)$-plane, $\phi_2$ in the $(4,5)$-plane,
and so on. Then the interaction potential is given by~\cite{pol} \be
V=-\int_0^\infty{{\rm d}t\over
t}(8\pi^2\alpha't)^{-1/2}\exp\left(-{ty^2\over 2\pi\alpha'}\right)
\prod_{a=1}^4{\vartheta_{11}(i\phi'_at/\pi,it)\over
\vartheta_{11}(i\phi_at/\pi,it)}~, \ee where $\phi'_a$ is a function
of the angles $(\phi_1, \dots, \phi_4)$.  Clearly the simplest case is
when just one angle, which we call $\phi$ is nonzero, meaning that the
branes are not aligned along just one direction. In this case, the
mass of the lightest excitation is~\cite{pol} \be m^2={y^2\over
4\pi^2\alpha^{'2}}-{\phi\over 2\pi\alpha'}~~ \mbox {with} ~~
0<\phi\leq \pi~.  \ee As the branes come closer, $y\ll\sqrt\alpha'$,
the lightest excitation mode becomes tachyonic, denoting an
instability. The two branes can lower their energy by reconnection,
which implies the process of unwinding. Since the two $D4$-branes are
already aligned in three directions (they intersect on a
$3=(4-1)$-dimensional sub-manifold) they can reconnect. This argument
can be generalised~\cite{dks} for the case of two intersecting
$D4$-branes at four arbitrary angles $0<\phi_i\leq \pi ~(\mbox{with}
~i=1,\cdots 4)$.

Within IIB string theory we do not have 4-dimensional branes; there
exist only odd dimensionality branes. The case of two interacting
$D5$-branes reduces to the case of two $D4$-branes. The reason for
this is that two $D5$-branes generically intersect along a line and
therefore applying $T$-duality along the intersecting direction we
obtain two $D4$-branes intersecting at a point.

Under the four hypotheses stated above, the following scenario has
been proposed~\cite{dks} to explain the origin of spacetime
dimensionality within IIB 10-dimensional string theory.  Consider an
ideal gas of $Dp$-branes embedded in a nine-dimensional bulk.  The
allowed dimensionality of the branes can be $p=1, 3, 5, 7, 9$.  Since
$D9$-branes fill the entire bulk, they overlap and thus they can
immediately reconnect and evaporate. Two $D7$-branes intersect on a
5-dimensional manifold, so they easily reconnect once they align one
more direction, and consequently unwind. The last ones to evaporate
are the $D5$-branes, since they intersect on a one-dimensional manifold,
thus  they have to align 3 directions before they can
reconnect. At the end the only branes surviving in the bulk are the
3-dimensional branes and $D$-string, together with closed string
modes.

This scenario remains valid~\cite{dks} if intersections between branes
of unequal dimensionality is allowed.  Applying the criterion for
brane intersections, one can easily check that $D3$-branes generically
only intersect with 7- and 9-dimensional branes, which are exactly the
ones to evaporate first. Thus, once $D7$- and $D9$-branes have
evaporated, $D3$-branes survive unaffected.

The criterion~\cite{dks}, Eq.~(\ref{bi}), for brane interactions has
been also used~\cite{kr} in the context of {\bf no} compact
dimensions. Consider~\cite{kr} a conventional, but higher-dimensional,
Friedmann-Lema\^itre-Roberston-Walker (FLRW) universe filled with
equal number of all possible branes and anti-branes.  Letting it to
expand it was argued~\cite{kr}, under some assumptions, that the only
consistent evolution is the one where the FLRW universe is dominated
by 3- and 7-dimensional branes. In this proposal the assumption of a
FLRW geometry is crucial, however it may not be well justified. In
addition, the dominant branes were found to be 3- but as well
7-dimensional.

\section{Conclusions}
We are living in a universe with three spatial dimensions and one is
therefore curious to understand the underlying reason.  If string
theory is indeed the correct fundamental theory, then the explanation
should lie within string theory itself. In addition, since string
theory requires a higher-dimensional space, either the theory can
explain the reduction of spatial dimensions or it is
inconsistent.

I have reviewed the scenarios proposed to explain the origin of
spacetime dimensionality. They can be divided into two classes,
Kaluza-Klein theories and braneworld models. Thus, I have reviewed the
string gas scenario and, a more recent proposal, based in brane
interactions within models with large extra dimensions. Even though it
is indeed difficult to actually prove the validity of these proposals,
one can see how they give indeed some hints on the origin of the
spacetime dimensionality.

\section*{Acknowledgements}
It is a pleasure to thank the organisers of the Conference
``Challenges in Particle Astrophysics'' - $6^{\rm th}$ {\sl Rencontres
du Vietnam}, for inviting me to present this work. I acknowledge
financial support from the Royal Society conference grants scheme
(Conference Grant 2006/R1).

\end{document}